\documentclass[12pt]{article}
\usepackage[letterpaper, portrait, top=1in, bottom=1in, left=1in, right=1in]{geometry}      
\usepackage{setspace}
\usepackage{subcaption}
\usepackage{graphicx}
\usepackage{amsmath}
\usepackage{amssymb}
\usepackage{epstopdf}
\usepackage{algorithm}
\usepackage{algpseudocode}
\usepackage{hyperref}
\usepackage{mdframed}
\usepackage{soul}
\usepackage{xcolor}
\usepackage[style=apa, backend=biber]{biblatex}
\newmdtheoremenv[innerrightmargin=22pt]{assumption}{Assumption}

\begin{document}
\begin{titlepage}
\title{Propensity score adjustment when errors in achievement measures inform treatment assignment}
\author{Joshua Wasserman\footnote{Department of Statistics, University of Michigan, Ann Arbor, MI, USA}, Ben B. Hansen\footnotemark[1]\hspace{5pt}, and Michael R. Elliott\footnote{Department of Biostatistics, University of Michigan, MI, USA}\hspace{5pt}\footnote{Survey Research Center, Institute for Social Research, University of Michigan, Ann Arbor, MI, USA}\hspace{5pt}\footnote{The research reported here was supported by the Institute of Education Sciences, U.S. Department of Education, through grants R305D210029 and R305S220002. The opinions expressed are those of the authors and do not represent views of the Institute or the U.S. Department of Education.}}
\clearpage\maketitle
\thispagestyle{empty}
\vspace{-14pt}
\begin{abstract}
U.S. state education agencies mark schools displaying achievement gaps between demographic subgroups as needing improvement. Some schools may have few students in these subgroups, such that average end-of-year test scores only noisily measure the average “true” score–the score one would expect if students took the test many times. This, in addition to the masking of small subgroup averages in publicly available assessment data, poses challenges for evaluating interventions aimed at closing achievement gaps. We introduce propensity score estimates designed to achieve balance on subgroup average true scores. These estimates are available even when noisy measurements are not and improve overlap compared to those that ignore measurement error, leading to greater bias reduction of matching estimators. We demonstrate our methods through simulation and an application to a statewide initiative in Texas for curbing summer learning loss.
\end{abstract}
Keywords: propensity score; matching; differential measurement error; learning loss; weighting
\end{titlepage}

\section{Less is more? To address or not address confounding measurement error in propensity scores}\label{intro}
U.S. federal law mandates that states have an accountability system for identifying schools in need of ``targeted support and improvement'', those where one or more particular demographic groups---``economically disadvantaged students, students in major racial and ethnic groups, children with disabilities, and English learners''---consistently underperform \parencite{essa}. Local education agencies (LEAs) must approve identified schools' plans for addressing these performance gaps. Texas's Additional Days School Year initiative (ADSY), a state-funded program aimed at reducing summer learning loss, offers one example of an intervention LEAs may require schools to implement as part of their corrective efforts.

A quasi-experimental evaluation of ADSY's efficacy might address LEAs selecting schools into the intervention by estimating a propensity score \parencite[PS;][]{rosenbaum_central_1983} for each school in the state, then matching or weighting schools that did not add days to their calendar such that the PS distribution is comparable to the PS distribution for schools that did. Given the relationship between subgroup achievement and selection into the intervention, the literature on PS estimation would suggest including measures of subgroup achievement in the PS model. However, the subgroups of interest may only comprise a mere handful of students at a given school. 

Achievement measures for small subgroups can be subject to two sources of variability. The first is year to year sampling variability of students in the subgroup. Comparing groups of ADSY and non-ADSY schools with similar distributions of subgroup achievement prior to the intervention provides a basis for estimating the effect of the initiative, but in a given year, the average score in a small subgroup may reflect students whose achievement particularly outpaces or lags behind their peers. Though academic performance for this group of students may deviate from what is typical for the subgroup, this is not of critical concern for PS estimation given that it is those students whose scores inform ADSY participation. Averages of student test scores are further subject to measurement error, however, which is the subject of our attention in this manuscript.

Due to high variability of performance on test day, the average scores in each subgroup may be a poor reflection of what one might expect those students to score if they took the test many times. That is, in view of classical test theory, which models a student's ``obtained'' score on a test as $W=X+\epsilon$, where $X$ denotes their average score over an infinite number of times taking the test (their ``true'' score) and $\epsilon$ represents a random test-day fluctuation \parencite{traub_understanding_1991, harville_standard_1991}, the average obtained score for students in a small subgroup may exhibit substantial variation around the students' average true score.

The fact that it is the average of obtained scores $W$, not true scores $X$, that factors into ADSY enrollment $T$ invokes dueling arguments in the PS estimation literature. On the one hand, some would take this fact to suggest fitting a PS model to the obtained scores $W$ without adjusting for $\epsilon$, the error in measuring the average true score $X$ \parencite{lockwood_matching_2016, hong_propensity_2019}. However, another section of the literature finds that including covariates in the PS model that correlate weakly with the outcome of interest $Y$ can harm both the bias and efficiency of effect estimates \parencite{westreich_role_2011, brookhart_variable_2006, rubin_estimating_1997}. If the test-day fluctuations $\epsilon$ have no correlation with post-intervention outcomes $Y$ conditional on true scores $X$, then $X$ will correlate more strongly with $Y$ than $W$ will. This suggests efforts should be made to form a comparison group of schools whose students have similar true scores to ADSY schools, not of schools whose students may have coincidentally over- or underperformed on test day prior to the intervention.

Balancing $e(W)$, a PS estimated from error-prone measurements, does not balance the error-free variable $X$ \parencite{raykov_propensity_2012, de_gil_how_2015}. Adjusting $W$ yields a valid balancing score in certain cases of ``nondifferential'' measurement error, that is, when $W$ and $T$ are conditionally independent given $X$ \parencite{lockwood_matching_2016}, but ADSY does not provide such a case. In fact, the correlation between $W$ and $T$ independent of $X$ substantially challenges the prospects of balancing intervention and non-intervention schools on $X$. \textcite{lockwood_matching_2016} identified the development of such balancing methodology as a key problem for future research.

We introduce two new methods for estimating propensity scores that balance schools on subgroups' average true scores rather than average obtained scores. The first uses regression calibration (RC), a technique that replaces $W$ in the PS model with estimates of $X$ \parencite{carroll_measurement_2006}. We estimate subgroup average true scores using empirical Bayes (EB) predictions from a hierarchical model of a type frequently applied in education research \parencite{raudenbush_hierarchical_2001}, but which also has connections to small area estimation in the survey literature \parencite{pfeffermann_small_2002}. Although the theoretical justification for RC estimates assumes nondifferential measurement error, the approach has gained popularity in no small part due to its demonstrated performance in a variety of modeling scenarios \parencite{rosner_correction_1989, carroll_measurement_2006}.

Our second approach jointly models the PS and the obtained scores conditional on the true scores and any other available covariates measured without error $Z$. Assuming $W$ captures the full effects of $X$ on $T$ (in other words, $X$ and $T$ are conditionally independent given $W$ and $Z$), the resulting PS model may be estimated from observed data by maximum likelihood (ML). To obtain a PS that conditions only on $X$ and $Z$, we marginalize $W$ out of the joint model. The required integral is analytically intractable, but we present a fast and accurate approximation to it using a device of \textcite{monahan_normal_1992}.

The ML PS estimates have less variability than PS estimates that condition on the error-prone variables, which has the important benefit of improving PS overlap between intervention and control groups. Certain matching techniques, such as matching within calipers, may exclude members of the intervention group for which no control units have comparable propensity scores, biasing the distribution of the covariates in the matched sample at the expense of maintaining balance \parencite{rosenbaum_matching_1985}, and possibly reducing the precision of subsequent effect estimates. Selecting a PS model may be driven by the tradeoff between confounder bias---the bias from making comparisons between groups that remain imbalanced in confounders---and bias from positivity violations---the bias from estimating intervention effects only from areas of overlap \parencite{westreich_invited_2010}. By reducing violations of positivity that occur as a consequence of poor PS overlap, and by controlling only for the part of the obtained scores that confound with post-intervention outcomes, ML PS estimates offer an enticing solution to this negotiation.

Our methods can be applied to aggregated student-level or public access school-level data, furnishing PS estimates in the latter case notwithstanding missing data issues associated with publicly available assessment data. Despite recent support for the use of school-level data for evaluating the effects of school-level interventions \parencite{munk_using_2021, jacob_assessing_2014, bloom_using_2007}, a key feature of publicly available school-level test scores is that states only report aggregates for subgroups of a minimum size, which in some states can be as many as forty students \parencite{jacob_assessing_2014}. By using predicted true scores instead of obtained scores, the RC and ML methods produce PS estimates even for those schools for which some subgroup averages are withheld. We find in our application to ADSY that the ML PS estimates yield better balance on the predicted true scores when applied to publicly available data with missing subgroup averages than PS estimates obtained from a logistic regression on the obtained scores in a dataset that is not missing those averages.

Section \ref{methods} defines notation and introduces the RC PS estimator, Section \ref{mle} introduces the ML estimator, Section \ref{simstudy} presents a simulation study that compares the methods to an estimator that ignores measurement error, Section \ref{app} applies them to ADSY, and Section \ref{discussion} concludes.
    
\section{An existing method for propensity score estimation with error-prone confounders}\label{methods}
\subsection{Notation}\label{notation}
We write that in the year prior to intervention, $n_{ijk}$ students at school $i$ in subgroup $k$ take end-of-year assessment $j$, obtaining an average scaled score of $W_{ijk}$. Maintaining the earlier test theory model, we write $W_{ijk}=X_{ijk}+\epsilon_{ijk}$, where $X_{ijk}$ denotes the average true score for those students on that year's assessment and $\epsilon_{ijk}$ denotes the average difference between students' true and obtained scores. We assume student-level differences between true and obtained scores have mean zero and are conditionally independent given the subgroup average true score $X_{ijk}$. Conditional on $X_{ijk}$, the subgroup average obtained score $W_{ijk}$ then has mean $\mathbb{E}[W_{ijk}|X_{ijk}]=X_{ijk}$ and variance $V(W_{ijk}|X_{ijk})=n_{ijk}^{-2}\sum_{\ell=1}^{n_{ijk}}\sigma^{2}_{ijk\ell}$, where $\sigma^{2}_{ijk\ell}$ is the conditional variance of the student-level difference, which may be heterogeneous across students, given $X_{ijk}$.

At the school level, school $i$ has a vector of obtained scores $\textbf{W}_{i}$, a vector of true scores $\textbf{X}_{i}$, and a vector of testtaker counts $\textbf{n}_{i}$ for all subgroups of interest on all assessments. Given $\textbf{X}_{i}$, the obtained scores $\textbf{W}_{i}$ have conditional mean $\textbf{X}_{i}$ and conditional variance $\Sigma_{i}$, where $\Sigma_{i}$ encodes the conditional variances and covariances of the subgroup average random deviations $\epsilon_{ijk}$. Although deviations for students in the same subgroup on the same assessment are assumed conditionally independent given their subgroup's average true score, subgroups of interest may not be mutually exclusive, in which case a student's random deviation may contribute to both $\epsilon_{ijk}$ and $\epsilon_{ijk'}$ for $k\neq k'$. We discuss implications for estimating $\Sigma_{i}$ in Section \ref{rc}.

Intervention assignment is denoted by the binary variable $T_{i}$, which takes a value of 1 for schools in the intervention group and 0 for those in the control group. We represent $P(T_{i}=1|\cdot)$, the propensity score given a set of variables, by $e(\cdot)$. Measurement error-free confounders with intervention assignment are collected in the vector $\textbf{Z}_{i}$.

Under the stable unit treatment value assumption \parencite{rubin_comment_1980}, we write the average post-intervention obtained score in a subgroup as $Y_{ijk}=T_{i}Y_{ijk}(1)+(1-T_{i})Y_{ijk}(0)$ \parencite{splawa-neyman_application_1923, rubin_estimating_1974}, where $Y_{ijk}(1)$ and $Y_{ijk}(0)$ denote potential outcomes under intervention and control, respectively. The PS itself does not involve these outcomes, but they figure into PS estimation's goals and assumptions. The ultimate target of estimation is often an intervention effect average, which, under Assumption \ref{assump:ignorability}, we write as $\tau_{g}=\mathbb{E}[g(\textbf{Z}_{i}, \textbf{W}_{i}, \textbf{X}_{i})(Y_{ijk}(1)-Y_{ijk}(0))]$, where $g$ targets effects in a specific population by ``tilting'' the density of the school-level characteristics towards a targeted population \parencite{li_propensity_2019}.

\begin{assumption}\label{assump:ignorability}
Subgroup-level potential outcomes are conditionally independent of intervention assignment given average pre-intervention obtained scores, subgroup sizes, and error-free covariates. Formally:
    \begin{align*}
        Y_{ijk}(1), Y_{ijk}(0)&\hspace{1pt}\perp \hspace{1pt}T_{i}\hspace{3pt}|\hspace{3pt}\textbf{W}_{i}, \textbf{n}_{i}, \textbf{Z}_{i}
    \end{align*}
\end{assumption}

\subsection{Regression calibration}\label{rc}
A predominant method for addressing measurement error, RC uses the regression of $X_{ijk}$ on $W_{ijk}$ and $\textbf{Z}_{i}$ as a regressor instead of the error-prone $W_{ijk}$. For linear errors-in-variables models with nondifferential measurement error, this modification (in tandem with the usual linear model assumptions) yields unbiased estimates of the regression that would be obtained using the error-free $X_{ijk}$ \parencite{carroll_measurement_2006}. Consistent estimation of nonlinear models may be achieved in some cases with nondifferential measurement error \parencite{rosner_correction_1989, gleser_improvements_1990}. In many others, RC provides a highly suitable approximation \parencite{carroll_measurement_2006}.

The method relies on the availability of either a dataset with all three of $\textbf{X}_{i}$, $\textbf{W}_{i}$ and $\textbf{Z}_{i}$; a dataset with replicates of $\textbf{W}_{i}$ to allow for estimation of $\Sigma_{i}$; or an external estimate of $\Sigma_{i}$. The true scores $\textbf{X}_{i}$ are not observed in the case of state standardized testing, and most students do not retake an assessment multiple times in a given year testing cycle, but an external estimate of $\Sigma_{i}$ can be fashioned from state education agencies' own estimates of test score variability.

In line with recommendations from the Standards for Educational and Psychological Testing \parencite{standards_2014}, many agencies publish conditional standard errors of measurement (CSEMs). CSEMs estimate the variability of test-day fluctuations on an assessment for a student with a particular true score. We use the CSEM associated with the subgroup average true score $X_{ijk}$ to estimate $\sigma^{2}_{ijk\ell}$ for each student $\ell$ in the subgroup. Since agencies also report the number of students that take each end-of-year assessment by subgroup (even for those subgroups falling below the threshold for reporting scores), we may then estimate $V(\epsilon_{ijk}|X_{ijk})$ by $n_{ijk}^{-1}\hat{\sigma}^{2}_{ijk}$, where $\hat{\sigma}_{ijk}$ denotes the CSEM. Each element on the diagonal of $\Sigma_{i}$ may be estimated in this way.

As mentioned previously, subgroups whose scores feature in $\textbf{W}_{i}$ and $\textbf{X}_{i}$ may not be mutually exclusive, which introduces correlation between the elements of $\textbf{W}_{i}$. Publicly available testtaker counts by subgroup may not be disaggregated along multiple demographic variables, however; counts are often available by race/ethnicity and sex separately, for example, but they are not available for the intersections of race/ethnicity groups and sexes. The development that follows focuses on the case of mutually exclusive subgroups where these complications do not arise. Generalization to the case of non-exclusive subgroups with known overlap counts is immediate, but in cases with unknown extent of overlap, additional assumptions would be needed to identify $\Sigma_{i}$.

This external estimate of $\Sigma_{i}$ plays a second crucial role: it allows for EB predictions of $\textbf{X}_{i}$ from a two-level hierarchical linear model (HLM) that accounts for variability between schools at one level and variability between subgroups within schools at another. The model, shown in (\ref{eqns:hlm}), includes fixed effects for $\textbf{Z}_{i}$, a random intercept $\delta^{(c)}_{ij}\sim\mathcal{N}(0, \tau_{1}^{2})$ for each school $i$ on each grade-level assessment $j$, and a random intercept $\delta^{(s)}_{ijk}\sim\mathcal{N}(0, \tau_{2}^{2})$ for each subgroup $k$ at school $i$ on grade-level assessment $j$. Without the external estimate of $\Sigma_{i}$, $\tau_{2}$ would not be identifiable, since $\epsilon_{ijk}$ and the random effect $\delta^{(s)}_{ijk}$ both occur at the subgroup level.

\begin{equation}
    W_{ijk} = \gamma_{0} + \textbf{Z}_{i}\gamma_{z} + \delta_{ij}^{(c)} + \delta_{ijk}^{(s)} + \epsilon_{ijk} \label{eqns:hlm}
\end{equation}

The EB prediction $\widehat{X}_{ijk}=\hat{\gamma}_{0}+\textbf{Z}_{i}\hat{\gamma}_{z}+\hat{\delta}_{ij}^{(c)}+\hat{\delta}_{ijk}^{(s)}$ obtained from the HLM for each subgroup and assessment is included in a vector $\widehat{\textbf{X}}_{i}$ that features in the RC PS model, shown in (\ref{eqns:rc}), along with the error-free covariates $\textbf{Z}_{i}$. RC is simple to implement, so it is of interest to understand how well the PS estimates from (\ref{eqns:rc}) approximate $e(\textbf{X}_{i}, \textbf{n}_{i}, \textbf{Z}_{i})$, despite the absence of nondifferential measurement error.

\begin{equation}
    \log\frac{e(\widehat{\textbf{X}}_{i}, \textbf{n}_{i}, \textbf{Z}_{i})}{1-e(\widehat{\textbf{X}}_{i}, \textbf{n}_{i}, \textbf{Z}_{i})} = \beta_{0} + \widehat{\textbf{X}}_{i}\beta_{\widehat{X}} + \textbf{Z}_{i}\beta_{z} \label{eqns:rc}
\end{equation}

\section{\texorpdfstring{A novel approach to propensity score estimation \\ with confounding measurement error}{A novel approach to propensity score estimation with confounding measurement error}}\label{mle}
Theoretical justification for RC stipulates that measurement error is nondifferential, or formally:
\begin{equation}
    T_{i}\hspace{2pt}\perp\hspace{2pt}\textbf{W}_{i}\hspace{3pt} |\hspace{3pt} \textbf{X}_{i}\hspace{1pt}, \hspace{2pt}\textbf{Z}_{i} \quad\quad .
\end{equation}
In the case of school-level interventions, this means the obtained scores offer no information about intervention assignment outside of the information carried in the true scores. However, the Every Student Succeeds Act mandates that statewide accountability systems for identifying schools in need of targeted support and improvement account in some part for performance on annual assessments \parencite{essa}. This means initiatives in the ensuing improvement plan are undertaken in response to the obtained scores whether those scores reflect true scores or not. In the same way, organizations may try interventions based on summaries of error-prone survey responses from their employees, just as doctors may prescribe treatment from imperfect assessments of their patients' health. In such cases, those where Assumption \ref{assump:ignorability} captures the assignment mechanism, the PS model ought to reflect the role measurement error plays in the process.

Assumption \ref{assump:ignorability} is not standard in settings with measurement error. Unlike the far more common assumption of nondifferential measurement error, the obtained scores may still carry information about intervention assignment given the true scores, making them ``weak surrogates'' \parencite{lockwood_matching_2016}. The distinction between nondifferential measurement error (or ``strong surrogacy'' of the obtained scores) and Assumption \ref{assump:ignorability} leads to the different approaches developed in Sections \ref{rc} and \ref{mle}. 

A crucial consequence of Assumption \ref{assump:ignorability} is that $e(\textbf{W}_{i}, \textbf{X}_{i}, \textbf{n}_{i}, \textbf{Z}_{i})=e(\textbf{W}_{i}, \textbf{n}_{i}, \textbf{Z}_{i})$. By modeling $T_{i}|\textbf{W}_{i}, \textbf{X}_{i}, \textbf{n}_{i}, \textbf{Z}_{i}$ using a logistic distribution, and by modeling $\textbf{W}_{i}|\textbf{X}_{i}, \textbf{n}_{i}, \textbf{Z}_{i}$ as multivariate normal, $e(\textbf{X}_{i}, \textbf{n}_{i}, \textbf{Z}_{i})$ may be obtained by integrating out $\textbf{W}_{i}$ (with respect to base measure $G$) from the joint distribution of $T_{i}$ and $\textbf{W}_{i}$ given $\textbf{X}_{i}$, $\textbf{n}_{i}$, and $\textbf{Z}_{i}$. This joint modeling approach was explored in \textcite{lockwood_using_2020} as a way of estimating $\mathbb{E}[\textbf{X}_{i}|T_{i}, \textbf{W}_{i}, \textbf{n}_{i}, \textbf{Z}_{i}]$ in the context of regression calibration. Our strategy is to use this as a way of obtaining a direct expression for the desired PS estimate.

With $\phi$ denoting the standard normal density function, the integral of the joint distribution over the conditional distribution of $\textbf{W}_{i}|\textbf{X}_{i}, \textbf{n}_{i}, \textbf{Z}_{i}$ is given by:
\begin{equation}
    e(\textbf{X}_{i}, \textbf{n}_{i}, \textbf{Z}_{i}) = \int\big{(}1+\exp(-\beta_{0}-w\beta_{w}-\textbf{Z}_{i}\beta_{z})\big{)}^{-1}\big{(}\beta_{w}^{T}\Sigma_{i}\beta_{w}\big{)}^{-1}\phi(\frac{w\beta_{w}-\textbf{X}_{i}\beta_{w}}{\beta_{w}^{T}\Sigma_{i}\beta_{w}})dG(w\beta_{w}) \hspace{4pt} . \label{eqns:mle}
\end{equation}

The logistic-normal integral in (\ref{eqns:mle}) arises both in the context of logistic regression with Gaussian error-prone variables and in the context of logistic regression with random effects. Various approximations have been proposed to address its analytical intractability \parencite[e.g.,][]{crouch_evaluation_1990, carroll_on_1984, stefanski_covariate_1985, fuller_measurement_1987}. We apply the approximation of \textcite{monahan_normal_1992}, which represents (\ref{eqns:mle}) by a location-scale mixture of normal distribution functions. This approximation, shown in (5) for three mixture components, can be computed quickly and cheaply given its dependence only on software for standard normal distribution functions (see the Appendix for a table with the scaling constants $\{s_{t}:t=1, ..., 3\}$ and mixing probabilities $\{p_{t}:t=1, ..., 3\}$). Given that under Assumption 1 an ordinary maximum likelihood fit of $T$'s logistic regression on $W$ and $Z$ furnishes estimates of $\beta_{0}$, $\beta_{w}$, and $\beta_{z}$, we refer to such estimates---again using the HLM from (1) to generate EB predictions of $\textbf{X}_{i}$ and the external estimate of $\Sigma_{i}$ from Section 2.2---as ML PS estimates.

\begin{equation}
    e(\textbf{X}_{i}, \textbf{n}_{i}, \textbf{Z}_{i}) \approx \sum_{t=1}^{3}p_{t}\cdot\Phi(\frac{s_{t}(\beta_{0}+\textbf{X}_{i}\beta_{w}+\textbf{Z}_{i}\beta_{z})}{\sqrt{1+s_{t}^{2}\beta_{w}^{T}\Sigma_{i}\beta_{w}}}) \quad .\label{eqns:approx}
\end{equation}

Our literature review does not find previous application of the approximation in (\ref{eqns:approx}) to education data, but \textcite{monahan_normal_1992} observes empirically that when $\log(e(\textbf{W}_{i}, \textbf{X}_{i},$ $\textbf{n}_{i}, \textbf{Z}_{i}))$ is Gaussian with variance on the scale of our simulations in Section \ref{simstudy}, this approximation produces as accurate or even more accurate approximations than 20-point Gauss-Hermite quadrature. Furthermore, we find almost no difference from adding more mixture components in our application in Section \ref{app}: the 95th percentile of absolute differences between PS estimates approximated using three components and PS estimates approximated using five components on the logit scale is just 0.03.

\section{Simulation Study}\label{simstudy}
\subsection{Setup}
We assess the estimators from Sections \ref{rc} and \ref{mle} in a simulated setting where 500 schools elect to participate in intervention under Assumption \ref{assump:ignorability}. Each school $i$ has error-free covariates $\textbf{Z}_{i}$, a vector of 6 independent $\text{Beta}(1.5, 0.5)$ random variables that simulate proportions of students in demographic groups formed by binary categorizations. These covariates correlate weakly with the true subgroup average scores on a single test instrument for four non-overlapping groups of students. More concretely (and dropping the subscript $j$ given the focus on a single test), $X_{ik}=\gamma_{0}+\textbf{Z}_{i}\gamma_{z}+\delta_{i}^{(c)}+\delta_{ik}^{(s)}$ for $k=1, ..., 4$, where $\delta_{i}^{(c)}\sim\mathcal{N}(0, \tau^{2}_{1})$, $\delta_{ik}^{(s)}\sim\mathcal{N}(0,\tau^{2}_{2})$, and $\text{V}(\textbf{Z}_{i}\gamma_{z})/\text{V}(X_{ik})\approx0.1$. Previous work that decomposed the variation of achievement between students and schools \parencite{hedges_intraclass_2007, hedges_intraclass_2013} informs our choices of $\tau_{1}$, $\tau_{2}$, and the signal-to-noise-ratio (see the Appendix for exact values of our simulation hyperparameters).

We generate the average obtained score in each subgroup as $W_{ik}=X_{ik}+\epsilon_{ik}$, where $\epsilon_{ik}|n_{ik}\sim\mathcal{N}(0, n^{-1}_{ik}\sigma^{2})$ and $\epsilon_{ik}\perp X_{ik}$. Unlike the CSEMs reported by state agencies, the CSEM in our simulations is constant across values of $X_{ik}$, hence our use of $\sigma^{2}$ in lieu of $\sigma^{2}_{ik}$. To assess the robustness of PS estimators to varying degrees of measurement error, we simulate the subgroup sizes in two ways. One approach simulates four ``large'' subgroups, where the number of students in each group is simulated from a $\text{Uniform}\{1, 120\}$ distribution. The other simulates two subgroups with a ``moderate'' number of students, simulating their sizes independently from a categorical distribution over $\{1, 2, \dots, 120\}$ with a mean of 44, and two ``small'' subgroups, simulating independently from a categorical distribution over $\{1, 2, \dots 60\}$ with a mean of 6. We derive $\sigma$ based on the work that informs our other hyperparameters. For the average ``large'' subgroup, this yields $V(\epsilon_{ik}|n_{ik})=32.3$; for the average ``moderate'' subgroup, $V(\epsilon_{ik}|n_{ik})=37.7$; and for the average ``small'' subgroup, $V(\epsilon_{ik}|n_{ik})=99.6$.

The probability of intervention is given by $P(T_{i}=1|\textbf{W}_{i}, \textbf{X}_{i}, \textbf{n}_{i}, \textbf{Z}_{i})=(1+\exp(-\beta_{0}-(\textbf{W}_{i}-\overline{\textbf{W}})\beta_{w}-(\textbf{Z}_{i}-\overline{\textbf{Z}})\beta_{z}))^{-1}$, where the chosen coefficients (provided in the Appendix) result in a marginal probability of intervention of about 1/3. Under the chosen assignment mechanism, for all $k=1, ..., 4$, intervention schools have values of $X_{ik}$ approximately 0.7 standard deviations lower than non-intervention schools.

Post-intervention potential outcomes are given by $Y_{ik}(T_{i})=X_{ik}+T_{i}\cdot g(\textbf{Z}_{i}, X_{ik})+\xi_{ik}$, where $\xi_{ik}|n_{ik}\sim\mathcal{N}(0, n^{-1}_{ik}\sigma^{2})$ and $\xi_{ik}\perp T_{i}, \textbf{Z}_{i}, X_{i1}, ..., X_{i4}$. Our choice of $g$, an exponentially decreasing function of $X_{ik}/1000$, results in a population average effect of intervention on intervention schools (abbreviated as ``ETT") of 5.5 points. This approach to outcome generation is somewhat of a simplification in that pre-intervention true scores averages $X_{ik}$ generally will not coincide with post intervention-averages $\mathbb{E}[Y_{ik}(0)|X_{ik}]$ due to student growth. Importantly, however, post-intervention potential outcomes are made to depend on $X_{ik}$ but not $W_{ik}$, despite the statistician having non-direct access to $X_{ik}$.

Following data generation, we use the RC approach from Section \ref{rc} (using the true $\sigma$ to form the external estimate of $\Sigma$), the ML approach from Section \ref{mle} (also using the true $\sigma$ to estimate $\Sigma$), and an ordinary ``naive'' logistic regression of $T_{i}$ on $\textbf{W}_{i}$ and $\textbf{Z}_{i}$ to produce three different sets of PS estimates. Each set informs a separate routine of full matching within calipers \parencite{hansen_optimal_2006}. This approach to matching retains any intervention school with a PS within one caliper width of a control school for analysis. Each retained school forms a part of a single matched set (potentially with other retained intervention schools) that we restrict to having no more than five control schools per intervention school.

Matching within calipers brings the issue of positivity to the fore: the procedure retains more intervention schools for analysis when the PS distributions for intervention and control schools have a greater region of overlap. We run simulations using calipers with widths of 0.5, 0.7, and 1 logits, assessing these approaches based on the percentage of intervention schools that remain unmatched across simulations. 

Once the three matched samples have been obtained, we compute standardized differences in weighted means of $X_{ik}$, denoted $d_{s}(X_{ik})$, and unstandardized differences in weighted means of $Y_{ik}$, denoted $d(Y_{ik})$, between intervention and non-intervention schools in each sample. Intervention schools receive a weight of 1 in the weighted means, while control schools are weighted by the odds of intervention in their matched set: for example, 1 for a control school paired to a single intervention school; 1/2 for either control school within a 1:2 matched set; and 2 for the control school falling in a 2:1 matched set. We average $d_{s}(X_{ik})$ across true scores for large, moderate, and small subgroups and compare these averages both across samples and to the standardized differences in unweighted means in the unmatched sample.

The differences $d(Y_{ik})$ figure as estimates of the ETT, so we take their average signed and root-mean-squared distance from the population ETT to compute the bias and root-mean-squared error (RMSE) of the matching estimators. We also compute these metrics for an estimator that weights control schools by the marginal odds of intervention in the unmatched sample, estimators that use the three sets of PS estimates as Horvitz-Thompson-type weights targeting the ETT \parencite{horvitz_generalization_1952}, and estimators that average differences between observed outcomes for intervention schools and predicted outcomes under the counterfactual from a regression on a penalized spline of the PS estimate in the control group \parencite[PENCOMP;][]{zhou_penalized_2019}.

\subsection{Results}
Table \ref{tables:unmatched_pct} reports the average percentage of intervention schools that failed to match in our simulations. Regardless of caliper width, ML and RC PS estimates led to a greater number of intervention schools matching than PS estimates using the naive approach. When subgroups had larger numbers of students, matching on ML estimates within calipers of 1 logit left 7\% of intervention schools unmatched, matching on RC estimates left 8\% unmatched, and matching on naive estimates left 10\% unmatched. When subgroups had moderate or small numbers of students (and were prone to more substantial measurement error), the improvement was striking: matching on naive PS estimates within a caliper of 1 logit excluded 21\% of schools in the intervention group from analysis; matching on RC estimates excluded 5\%, and matching on ML estimates excluded just 2\%.

\begin{figure}[!t]
\centering
\includegraphics[width=\textwidth,height=0.3\textheight]{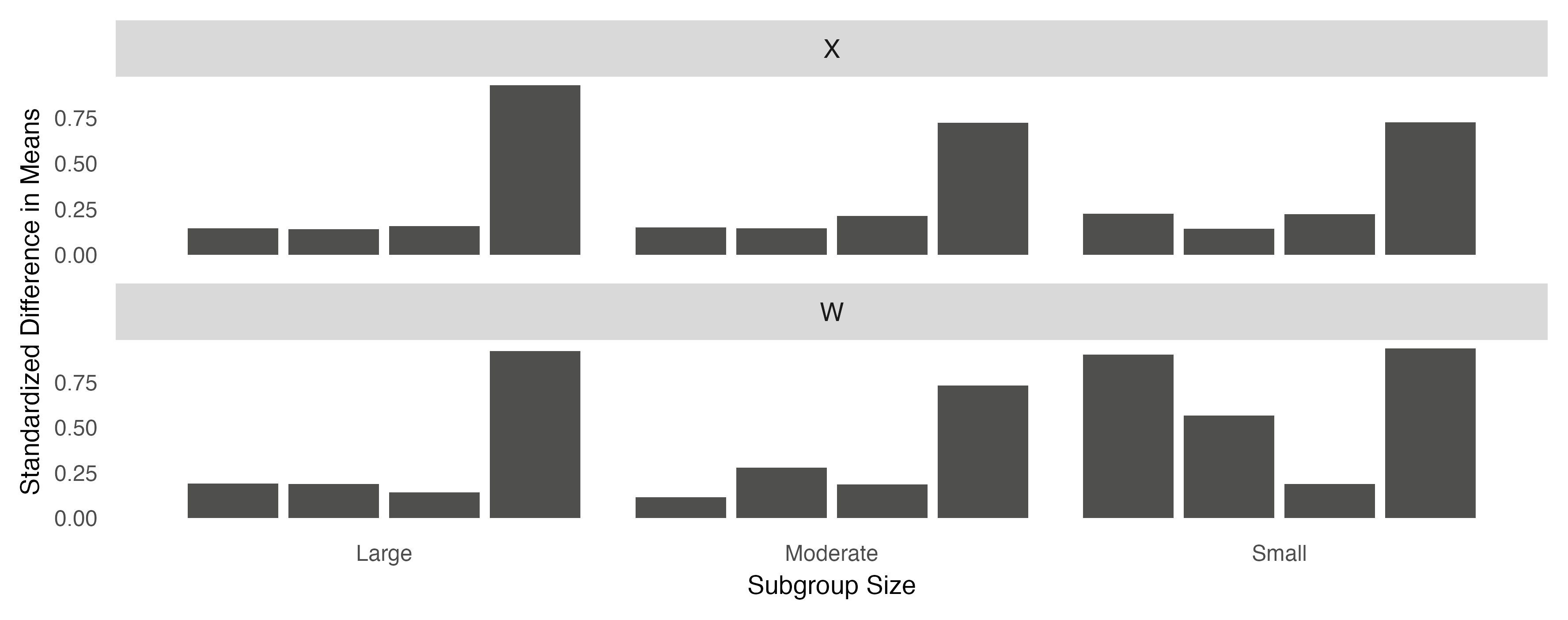}
\caption{Average standardized differences in weighted means of subgroup true scores ($X$) and obtained scores ($W$). Notes. Bars at far right of each group of four illustrate differences between groups in the unmatched sample, while those at mid-right, mid-left, and far left illustrate differences between groups in matched samples obtained from matching on naive, regression calibration (RC), and maximum likelihood (ML) PS estimates, respectively. Bars average across subgroups whose sizes are simulated from the same distribution.}
\label{figures:XWbalance}
\end{figure}

Figure \ref{figures:XWbalance} displays imbalances in $\textbf{X}_{i}$ and $\textbf{W}_{i}$ between intervention and non-intervention schools. Without matching, intervention and non-intervention schools differed drastically in $\textbf{X}_{i}$ and $\textbf{W}_{i}$. Matching on naive PS estimates improved balance, but matching on ML or RC estimates further reduced imbalance in $\textbf{X}_{i}$ in large and moderate subgroups. Matching on ML estimates produced a matched sample with roughly equal balance on small subgroup true scores compared to the matched sample produced by matching on naive estimates, while matching on RC PS estimates produced a more balanced matched sample than both. In small subgroups, the matching routines that used ML or RC PS estimates hardly improved balance on $\textbf{W}_{i}$ at all, indicating these approaches expend balance on $\textbf{W}_{i}$ to achieve better balance on $\textbf{X}_{i}$.

\begin{table}[!t]
\centering
\begin{tabular}{||c|c|c|c|c||}
\hline
Subgroup Sizes & Caliper Width & ML & RC & Naive \\
\hline
 & 0.5 & 11 & 12 & 16 \\
All large & 0.7 & 9 & 10 & 13 \\
 & 1 & 7 & 8 & 10 \\
\hline
& 0.5 & 4 & 8 & 28 \\
2 moderate, 2 small & 0.7 & 3 & 7 & 25 \\
& 1 & 2 & 5 & 21 \\
\hline
\end{tabular}
\caption{Average percentage of intervention schools for which full matching within calipers fails to find a match. Notes. Percentages reported for matching using maximum likelihood (ML), regression calibration (RC), and naive PS estimates. Averages taken over 1000 simulations. Caliper widths reported on the logit scale.}
\label{tables:unmatched_pct}
\end{table}

\begin{figure}[!t]
\centering
\includegraphics[width=\textwidth,height=0.3\textheight]{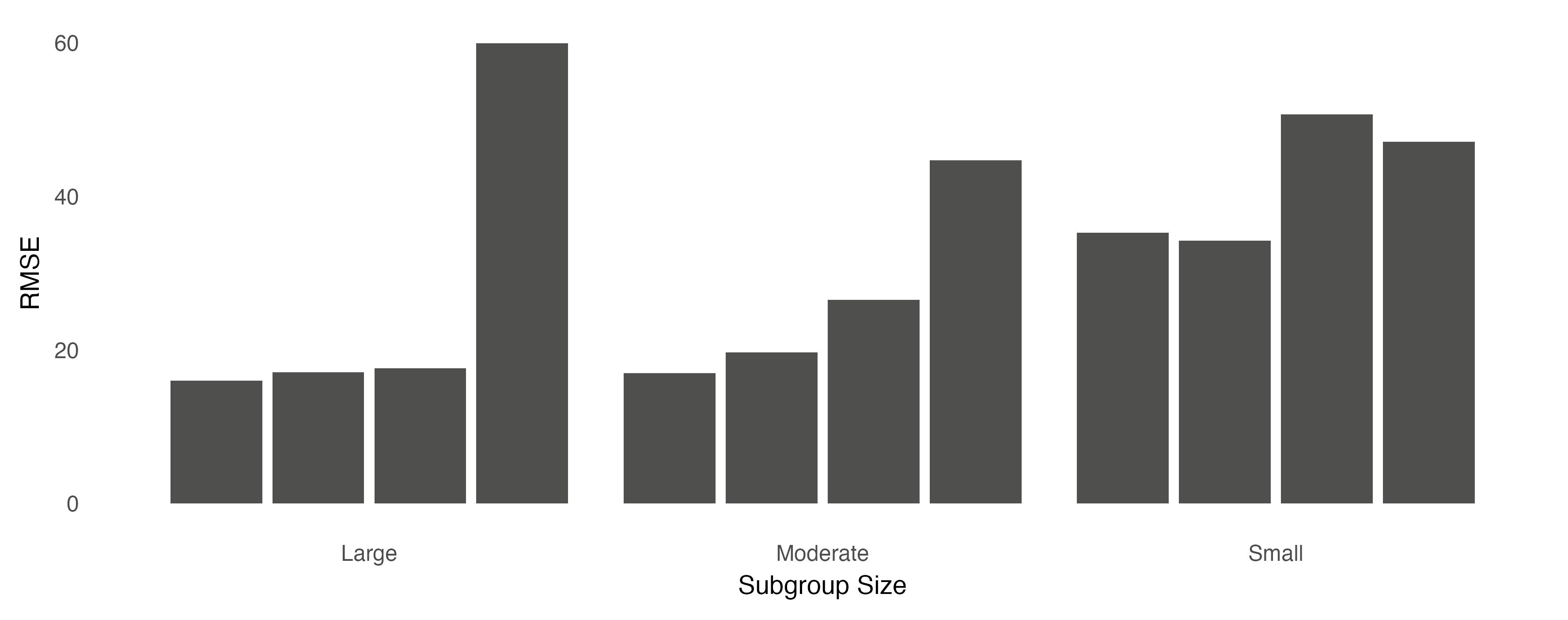}
\caption{Root-mean-squared error (RMSE) of matching estimators of the average effect of intervention on intervention schools (ETT). Notes. Bars at far right of each group of four show the error of the estimator weighted by the marginal odds of intervention participation in the unmatched sample, while those at mid-right, mid-left, and far left show the error of the stratum-weighted estimator applied to the matched samples obtained from matching on naive, regression calibration (RC), and maximum likelihood (ML) PS estimates, respectively. Bars average across subgroups whose sizes are simulated from the same distribution.}
\label{figures:Ybalance}
\end{figure}

Stratum-weighted estimates of the ETT demonstrated less error when applied to matched samples formed by matching on ML or RC PS estimates than those formed by matching on naive PS estimates. The matching estimator applied to the sample matched on ML estimates had an RMSE of 16 points when all subgroups had a large number of students, compared to 17.1 points in the sample matched on RC estimates, 17.6 points in the sample matched on naive estimates, and 60 points for the marginal odds-weighted estimator in the unmatched sample. As illustrated in Figure \ref{figures:Ybalance}, the improvement is slightly more pronounced in moderate subgroups, but unmistakable in small subgroups: in this latter case, the matching estimators applied to the samples matched on ML and RC estimates have RMSEs of 35.3 and 34.2 points, respectively, about a third less than the estimator applied to the sample matched on naive estimates.

Weighting and PENCOMP estimators exhibit the same trend. PENCOMP applied with either ML or RC PS estimates achieved the lowest RMSE in each subgroup size, as seen in Table \ref{tables:all_res}. For large subgroups, this minimum RMSE was 10.5, compared with 11.2 for the PENCOMP estimator with naive PS estimates; for moderate subgroups it was 13.4 compared with 16.8; and for small subgroups it was 24.4 compared with 31.6. The empirical decomposition of the bias and variance presented in Table \ref{tables:all_res} reveals that while estimators leveraging the ML estimates have greater bias, they have much less variance than estimators that make use of other PS estimates. Worth noting is that the bias of all three estimators may be reduced by adjusting for covariates. Finding hyperparameters for the matching procedure that optimize balance in confounders could further reduce the bias and RMSE of the matching estimators, while weight trimming could improve the performance of the weighting estimators.

\begin{table}[!t]
    \centering
    \begin{tabular}{||c|c|c|c|c|c|c|c|c||}
    \hline
     & & & \multicolumn{2}{|c|}{ML} & \multicolumn{2}{|c|}{RC} & \multicolumn{2}{|c|}{Naive} \\ \cline{4-9}
      Subgroup sizes & Subgroup & Method & Bias & RMSE & Bias & RMSE & Bias & RMSE\\
      \hline
        & Large & Matching & -1.4 & 16.0 & -1.3 & 17.1 & -0.6 & 17.6 \\
        All large & Large & Weighting & 8.5 & 13.2 & 9.6 & 19 & 10.2 & 19 \\
        & Large & PENCOMP & -1.4 & 10.5 & -0.4 & 11.3 & -0.1 & 11.2 \\
        \hline
        & Moderate & Matching & 8.2 & 17 & -4.4 & 19.7 & -0.7 & 26.5 \\
        & Moderate & Weighting & 5.8 & 9.9 & 10.7 & 19.6 & 16.7 & 29 \\
        2 moderate, & Moderate & PENCOMP & 9.1 & 14.5 & -1.6 & 13.4 & 0.2 & 16.8 \\
        \cline{2-9}
        2 small & Small & Matching & -12.9 & 35.3 & -0.9 & 34.2 & 0.4 & 50.7 \\
        & Small & Weighting & 20 & 25.9 & 24.2 & 35.8 & 35.9 & 53.9 \\
        & Small & PENCOMP & -12.2 & 26.2 & -0.3 & 24.4 & 0.5 & 31.6 \\
        \hline
    \end{tabular}
    \caption{Bias and root-mean-squared error (RMSE) of matching, weighting, and PENCOMP estimators of the average effect of intervention on intervention schools. Notes. Results for propensity scores estimated by maximum likelihood (ML), regression calibration (RC), and a logistic regression on observed scores (naive) are presented. The RMSE of the matching estimators are the same as those shown in Figure \ref{figures:Ybalance}.}
    \label{tables:all_res}
\end{table}

\subsection{Sensitivity to parametric assumptions}
This work makes two key parametric assumptions: first, that the CSEMs parameter values are known, and second, for the ML estimates, that the measurement error is Gaussian. We consider deviations from these assumptions and illustrate results in the Technical Appendix.

First, we consider CSEMs that either underestimate or overestimate the true variability of the measurement error by 25\%, a magnitude we believe would be far larger than anticipated in practice. Still, matching rates are similar for the large group sizes regardless of CSEM misestimation; for moderate/small group sizes, the failure to find full matching is somewhat higher in when CSEM is underestimated and somewhat lower when CSEM is overestimated.  The achieved RMSEs are also similar in the large groups and in the small groups when CSEM is overestimated, but somewhat larger when CSEM is overestimated.

Next, we consider a non-Gaussian (mixture of normal) measurement error distribution. Specifically, we find that by simulating $\epsilon_{ik}|n_{ik}\sim n_{ik}^{-1}(0.6\mathcal{N}(0, \sigma^{2})+0.4\mathcal{N}(-75,100^{2}))$ we introduce skew to the measurement error distribution. We simulate measurement error from this conditional distribution for pre-intervention outcomes, but for post-intervention outcomes we continue to simulate measurement error from a single Gaussian. Here we find the match rate unchanged from when the data is generated under the assumed model, with RSME similar when groups are large and somewhat increased when groups are small.

In sum, we find the results differ slightly from those reported here, particularly if the groups are small or the CSEM is underestimated, but the broad takeaways from comparing these three methods do not change.

\section{\texorpdfstring{Forming a matched comparison group for early \\ adopters of ADSY}{Forming a matched comparison group for early adopters of ADSY}}\label{app}
We now use our proposed methods to form a matched comparison group for schools that added days during the 2020-2021 academic year as part of ADSY, paying particular attention to pretest scores among Black and Hispanic students. That year, the median public school in Texas enrolled 233 students identifying as racially or ethnically Hispanic or Latino and 27 students identifying as Black or African-American \parencite{ccd}. Annual assessments were not taken at the end of the 2019-2020 academic year, however, due to the COVID-19 shutdown. The most recent available pretest being two years prior, PS estimates that leverage them may not eliminate confounding between ADSY and non-ADSY schools. Accordingly, we defer estimation of ADSY's effects to ongoing evaluations that also consider later student cohorts. Here, we instead provide results from a placebo test on average scores in 2019 from subgroups we did not use in PS estimation. Since there should be no difference between ADSY and matched non-ADSY schools in these scores, significant differences imply matching did not suitably remove selection bias.

Our PS model includes schools' charter, magnet, and Title I statuses as reported in the Common Core of Data \parencite[CCD;][]{ccd}; student body composition by sex, race/ethnicity, and socioeconomic status (also available from the CCD); and, as will be explained below, either averages of obtained scores or EB predictions of average true scores on 3rd, 4th, and 5th grade math and reading assessments for students in the two subgroups of interest. We obtain EB predictions from HLMs fit separately for each grade-level assessment, though we find no material difference in the resulting PS estimates when fitting one HLM across grades for math scores and one HLM across grades for  reading scores. Each HLM includes school and subgroup-within-school random effects, fixed effects for test language (either English or Spanish) and race/ethnicity, and the school status and demographic covariates from the PS model. Since the CSEMs that inform the external estimate of $\Sigma_{i}$ are conditional on true scores, we fit each HLM twice, first using the CSEMs associated with the subgroup's average obtained score, then, after obtaining EB predictions from the first fit, with the CSEMs associated with the EB predictions. Our final EB predictions are weighted averages over test language, with the proportion of students in a subgroup taking the language in English and Spanish serving as weights.

TEA withholds average scores from public-access assessment data for subgroups comprising fewer than five students. Through a data-sharing agreement with the agency, we have access to student-level data that we could use to impute the missing subgroup averages exactly. However, rather than apply our proposed methods to a complete dataset with these imputations, we apply them to the school-level data as provided to the public. Excluding schools from the PS and HLM model fits whose pre-test scores have been withheld does not preclude either the ML or RC estimator from producing PS estimates for those schools. The only necessary artifacts for ML PS estimates are the coefficient estimates from the fitted PS model, EB predictions of the average true scores, and estimates of standard errors of measurement, all of which are available for schools missing one or more subgroup averages. RC PS estimates are available for these schools as well, as the PS model is fit not to the subgroup averages but the EB predictions of the true scores.

We compare balance from matching on these PS estimates to the balance obtained by matching on PS estimates from a model fit to the noisy subgroup averages in a dataset that does impute the missing averages with student-level data. Without such imputations, the naive estimator could not confer PS estimates upon schools with any subgroup averages withheld from the public-access data. Still, by giving the naive estimator this boost and allowing a comparison, we can assess whether school-level matching benefits from the use of restricted-access student-level data in place of public-access school-level data.

\begin{figure}[!t]
\includegraphics[width=\textwidth,height=0.33\textheight]{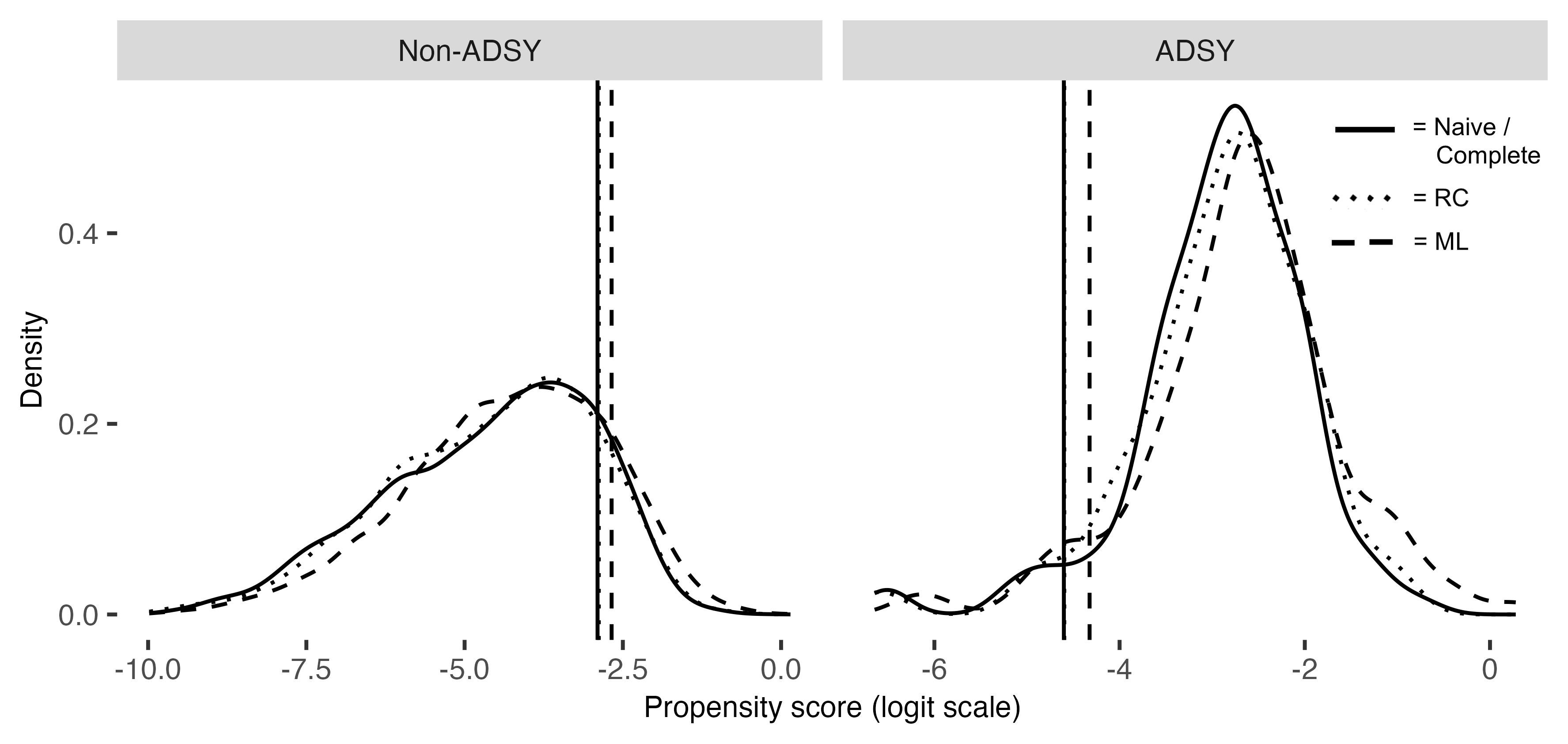}
\caption{Distributions of PS estimates on the logit scale among non-ADSY and ADSY schools. Notes. Vertical lines mark the mean PS estimate in each group.}
\label{figures:adsy_ps}
\end{figure}

Figure \ref{figures:adsy_ps} shows the distributions of the three PS estimates among ADSY and non-ADSY schools. Among both the intervention and control groups, the ML estimates are shifted away from the more extreme areas of the support, with bumps in the distribution shifted noticeably away from the lower end of support when compared to the naive and RC estimates. The mean ML estimate in both groups, given by the dashed vertical lines, is slightly higher than the mean naive and RC estimate given by the solid and dotted vertical lines, respectively. Table \ref{tables:ess} presents the number of ADSY schools retained in and the effective sample size of each matched sample. With a maximum ratio of 10 controls to 1 ADSY school in each matched set, we see that although each routine included all ADSY schools for analysis, matching on ML PS estimates achieves a slightly larger effective sample size.

\begin{table}[!t]
\centering
\begin{tabular}{||c|c|c|c||}
\hline
Metric & ML & RC & Naive/Complete \\
\hline
Matched ADSY schools & 125 & 125 & 125 \\
\hline
Effective sample size & 212.5 & 211 & 211 \\
\hline
\end{tabular}
\caption{Number of ADSY schools retained in matched samples and their effective sample sizes. Notes. Maximum likelihood (ML) and regression calibration (RC) PS estimates were generated without imputing withheld subgroup averages; naive PS estimates were generated using a dataset where withheld subgroup averages were imputed by averages computed from student-level data.}
\label{tables:ess}
\end{table}

Since the subgroup average true scores are unknown, Table \ref{tables:adsy_balance} reports standardized differences between ADSY and non-ADSY schools in EB predictions of those scores. Differences in the matched samples only figure matched sets where at least one ADSY and one non-ADSY school had students that took the assessment on which a comparison is being drawn; a matched pair is dropped from the comparison of Black or African American students on 5th grade math, for example, if either school did not have Black or African American students that took the 5th grade math assessment. Across subgroups and assessments, ML PS estimates achieve the best balance on average, with an average standardized difference between group means of 0.033 (presented in Table \ref{tables:adsy_balance} on a scale $10^{2}$ times the original scale). Matching on naive and RC PS estimates achieved slightly worse average balance, reporting average differences of 0.039 and 0.046, respectively. Matching on ML estimates also achieved the best balance on the scores for which the least balance was attained: the least balanced EB predictions had a standardized difference-in-means of 0.071, while matching on naive and RC estimates resulted in maximum differences of 0.091 and 0.193, respectively. However, after matching on RC estimates only two subgroups have residual covariate imbalances large enough to require post-matching covariate adjustment, according to the standards of the What Works Clearinghouse \parencite{wwc}, while matching on ML and naive estimates leaves three subgroups needing adjustment.

\begin{table}[!t]
\centering
\begin{tabular}{||c|c|c|c|c|c|c||}
\hline
Subgroup & Grade & Subject & ML & RC & Naive/ & Unmatched \\
& & & & & Complete & \\
\hline
Hispanic or Latino & 3 & M & 1.17 & 0.11 & 1.95 & 20.13 \\
Black or African-American & 3 & M & 4.91 & 16.26 & 4.05 & 9.10 \\
Hispanic or Latino & 3 & R & 0.06 & 2.94 & 0.69 & 2.94 \\
Black or African-American & 3 & R & 6.04 & 19.29 & 9.11 & 2.33 \\
Hispanic or Latino & 4 & M & 3.88 & 1.28 & 5.55 & 4.07 \\
Black or African-American & 4 & M & 3.40 & 1.10 & 1.08 & 9.00 \\
Hispanic or Latino & 4 & R & 1.18 & 0.73 & 3.18 & 3.37 \\
Black or African-American & 4 & R & 7.11 & 3.55 & 3.18 & 12.89 \\
Hispanic or Latino & 5 & M & 5.44 & 1.96 & 0.75 & 16.25 \\
Black or African-American & 5 & M & 0.71 & 0.84 & 2.79 & 2.02 \\
Hispanic or Latino & 5 & R & 4.67 & 4.60 & 7.64 & 11.59 \\
Black or African-American & 5 & R & 0.85 & 2.78 & 7.05 & 2.68 \\
\hline
Overall average & & & \textbf{3.28} & 4.62 & 3.92 & 8.03 \\
\hline
\end{tabular}
\caption{$10^{2}\hspace{3pt}\times $ standardized differences in weighted averages of empirical Bayes (EB) predictions of subgroup average true scores. Notes. Differences are presented for scores on 3rd, 4th, and 5th grade math (M) and reading (R) assessments. Means in the unmatched sample are unweighted, while means in the samples matched on maximum likelihood (ML), regression calibration (RC), and naive propensity score estimates are weighted by the odds in each matched set. The average difference across subgroup scores is reported in the bottom row, with the smallest average across samples given in boldface font.}
\label{tables:adsy_balance}
\end{table}

Table \ref{tables:adsy_estimates} reports the results of the placebo test. We use the estimator and associated variance estimator of \textcite{wasserman_dissertation_2026} to produce covariate-adjusted differences in four placebo outcomes: average math and reading scores for Asian 3rd graders, and average math and reading scores for White 3rd graders. The p-values in the table reflect a max-T multiplicity correction applied within families of hypotheses tested using the same set of PS estimates. Differences in math and reading scores for White 3rd grade students are not significant using any estimator, while differences for Asian students are significant at the 0.001 level for all estimators. The results across PS estimates do not noticeably differ, suggesting that balance in scores for Hispanic and Black students does not imply balance in scores for Asian students. One possible reason for the observed imbalance is exactly the issue we set out to address: in TEA's test score data from 2019, the median number of Asian students at a school taking the 3rd grade math assessment was exactly zero; the 75th percentile was only three. As we have discussed, subgroup averages from these small handfuls of students (when available at all) will be highly variable. In contrast, the median school had 13 white students take the 3rd grade math assessment, with the 75th percentile being 37 students. Balancing achievement for white students between ADSY and non-ADSY schools may be less of a challenge given these subgroup sizes.

\begin{table}[!t]
\centering
\begin{tabular}{||c|c|c|c|c|c||}
\hline
Subgroup & Subject & PS Estimator & Estimate (SE) & T-stat. & Adj. p-value \\
\hline
& & ML & 12.14 (8.25) & 1.47 & 0.45 \\
\cline{3-6}
White & M & RC & 14.19 (8.65) & 1.64 & 0.34 \\
\cline{3-6}
& & Naive & 12.67 (8.24) & 1.54 & 0.40 \\
\hline
& & ML & $42.78^{***}$ (11.55) & 3.71 & $8.58\cdot10^{-4}$ \\
\cline{3-6}
Asian & M & RC & $47.30^{**}$ (13.56) & 3.49 & $1.96\cdot10^{-3}$ \\
\cline{3-6}
& & Naive & $43.92^{**}$ (12.38) & 3.55 & $1.56\cdot10^{-3}$ \\
\hline
& & ML & 6.16 (7.09) & 0.87 & 0.85 \\
\cline{3-6}
White & R & RC & 8.08 (6.50) & 1.24 & 0.61 \\
\cline{3-6}
& & Naive & 2.10 (7.04) & 0.30 & 1.00 \\
\hline
& & ML & $41.38^{**}$ (12.75) & 3.24 & $4.73\cdot10^{-3}$ \\
\cline{3-6}
Asian & R & RC & $42.22^{**}$ (13.94) & 3.03 & $9.82\cdot10^{-3}$ \\
\cline{3-6}
& & Naive & $41.08^{**}$ (13.49) & 3.05 & $9.27\cdot10^{-3}$ \\
\hline
\end{tabular}
\caption{Results from a placebo test comparing test scores of white and Asian third graders in ADSY schools to their counterparts in matched control schools in the year prior to ADSY implementation. Notes. P-values are multiplicity-corrected using a max-T correction.}
\label{tables:adsy_estimates}
\end{table}

\section{Discussion}\label{discussion}
Evaluating school-level interventions implemented as part of targeted support and improvement plans requires addressing the noisy (and sometimes missing) test score averages that, in part, determine schools' participation. Our proposed PS estimators simultaneously address the error in these achievement measures as well as their possible missingness, demonstrating valuable improvement over the PS estimator that remedies neither.

First, they reduce the variability of the resulting PS estimates, leading to fewer violations of positivity that would necessitate the exclusion of intervention schools from analysis. Second, in many cases they better balance the expected test performance of students between intervention and matched control schools. These characteristics lead to substantial reductions in estimation error of subsequent effect estimators when compared to approaches that ignore measurement error. Although our proposed approaches had lower RMSE, our ML PS estimator was more biased when some subgroup average scores exhibited substantial measurement error but others did not. This contrasts with the setting where the measurement error for all four subgroup averages followed the same distribution. Future research is needed to understand how the ML estimator attends to subgroup averages with differing measurement error distributions.

We choose the HLM in (1) to obtain accurate predictions of $X_{ijk}$, but it may be replaced if the researcher prefers another model for predicting $X_{ijk}$. As detailed in Section \ref{intro}, given the ultimate task of PS estimation, we desire accurate estimation of the average true score associated specifically with the students in the subgroup in the year prior to intervention, this being the group of students whose scores determine treatment assignment. However, future work could explore the benefit of estimating year to year variability of subgroup average scores when longitudinal data is available. These estimates could be incorporated into the HLM for predicting average true scores or as a variance component and marginalized out as part of MLE PS estimation.

Our proposed methods may suffer if test score averages are withheld from publicly available data not at random; that is, non-reported averages are higher or lower on average than those reported in the data. The coefficient estimates from the PS model fit should reflect the conditional mean of intervention participation given noisy achievement measures and other covariates measured without error, while the EB predictions should reflect the mean achievement in the subgroup conditional on those error-free covariates. In general, many schools may not have enough students in subgroups of interest for those averages to be reported in school-level data, and this issue can be compounded when using averages from multiple grades in the PS model or when using data from states with even more restrictive reporting thresholds.

As noted previously, the availability of only marginal subgroup counts limits the ability to estimate correlations between measurement error of average obtained scores in overlapping subgroups. A future study might compare the performance of the ML approach when the PS model includes averages from overlapping subgroups while (a) $\Sigma_{i}$ is assumed diagonal, or (b) unreported subgroup sizes are imputed from the loglinear model of independence for the three-way contingency table of counts by school, subgroup 1 and subgroup 2. This may be an important investigation given previous discoveries that correlated measurement error can increase the bias of causal effect estimates \parencite{hong_propensity_2019}. Such extensions would still need an external estimate of the standard error of measurement, though, which could prove more difficult to accomplish in practice.

\setlength{\emergencystretch}{1em}
\printbibliography

\newpage
\section*{Appendix}
\setcounter{table}{0}
\renewcommand{\thetable}{A\arabic{table}}

\begin{table}[!h]
    \centering
    \begin{tabular}{||c|c|c||}
    \hline
      $t$ & $p_{t}$ & $s_{t}$ \\
      \hline
        1 & $0.252201578098282$ & $0.907930837449693$ \\
        2 & $0.585225059235736$ & $0.577787276140136$ \\
        3 & $0.162573362665982$ & $0.36403772947977$ \\
        \hline
    \end{tabular}
    \caption{The mixing probabilities, $p_{t}$, and scaling constants, $s_{t}$, associated with the approximation to the logistic-normal integral in (\ref{eqns:approx}). Notes. These are reproduced exactly from \textcite{monahan_normal_1992}.}
    \label{tables:approx_constants}
\end{table}

\begin{table}[!h]
    \centering
    \begin{tabular}{||c|c||}
    \hline
      Parameter & Value \\
      \hline
        $\tau_{1}$ & $40.229$ \\
        $\tau_{2}$ & $59.149$ \\
        $\sigma$ & $250$ \\
        $\beta_{0}$ & $-1.386$\\
        $\beta_{w, k}$ & $-0.0115$ \\
        $\beta_{z, k}$ & $0.05$ \\
        \hline
    \end{tabular}
    \caption{Parameter values used in the simulation study in Section \ref{simstudy}. The vector of coefficients $\beta_{w}$ in the manuscript has four elements, each equal to $\beta_{w, k}$. The vector of coefficients $\beta_{z}$ in the manuscript has six elements, the first two of which are $\beta_{z, k}$; the remaining coefficients are 0.}
    \label{tables:sim_params}
\end{table}

\end{document}